\documentclass[aps,prl,nofootinbib,twocolumn,superscriptaddress]{revtex4-1}
\usepackage{hyperref}
\hypersetup{colorlinks=true,linkcolor=blue,citecolor=red,urlcolor=blue}
\usepackage{graphicx}
\usepackage{xcolor}

\usepackage[normalem]{ulem}
\usepackage{adjustbox}

\usepackage{physics,amsmath,amsfonts,nicefrac,esint,amssymb,bbold}
\usepackage{array}
\usepackage{natbib}
\usepackage{graphicx}
\usepackage[absolute,overlay]{textpos}

\definecolor{mygreen}{HTML}{006400}

\newcommand{\non}{\nonumber\\}

\DeclareMathOperator{\sgn}{sgn}

\newcommand{\taup}{\tau^\prime}

\begin{document}

\title{Josephson wormhole in coupled superconducting Yukawa-SYK metals}%

\author{Aravindh S. Shankar}
\email{ashankar@ictp.it}
\affiliation{Instituut-Lorentz for Theoretical Physics, $\Delta$-ITP, Leiden University, The Netherlands.}
\affiliation{The Abdus Salam International Center for Theoretical Physics, Strada Costiera 11, 34151 Trieste, Italy}
 \author{Jasper Steenbergen}
 \affiliation{Instituut-Lorentz for Theoretical Physics, $\Delta$-ITP, Leiden University, The Netherlands.}
 %
\author{Stephan Plugge}
\affiliation{Instituut-Lorentz for Theoretical Physics, $\Delta$-ITP, Leiden University, The Netherlands.}
\affiliation{Silicon Quantum Computing Pty Ltd \& CQC2T, School of Physics, UNSW Sydney, Australia}
%
\author{Koenraad Schalm}
\email{kschalm@lorentz.leidenuniv.nl}
\affiliation{Instituut-Lorentz for Theoretical Physics, $\Delta$-ITP, Leiden University, The Netherlands.}

\begin{abstract}
\noindent
We show that two Yukawa-SYK models with a weak tunneling contact can have an exotic hybrid superconducting thermofield-double-like state that is holographically dual to a traversable wormhole connecting two black holes with charged scalar hair. The hybrid superconducting thermo-field-double/wormhole state is distinguishable by anomalous scaling of revival oscillations in the fermionic Green's function, but also in a unique Andreev-revival in the anomalous Green's function. The existence of this TFD/WH state surprisingly shows that the some quantum critical effects can survive the phase transition to superconductivity. This Andreev-revival is in principle an accessible signature of the transition to the TFD/WH phase detectable in the ac-Josephson current.
\end{abstract}

\maketitle

\noindent
The realization that the quantum critical, strongly correlated groundstate of the Sachdev-Ye-Kitaev (SYK) model
is in the same universality class as the groundstate of holographic models of charged anti-de-Sitter black holes has opened up a wide avenue to study exotic gravitational questions with quantum-mechanical Hamiltonians. One such question is the existence of traversable wormholes. Maldacena and Qi \cite{maldacena2018eternal}, based on an earlier work of Gao, Jafferis and Wall \cite{gaoTraversableWormholesDouble2017} and more recently others \cite{maldacenaDivingTraversableWormholes2017,maldacenaTraversableWormholesFour2020,cottrellHowBuildThermofield2019,bakBulkViewTeleportation2018,gaoRegenesisQuantumTraversable2019,fuTraversableAsymptoticallyFlat2019,bakExperimentalProbesTraversable2019,qiCoupledSYKModel2020a,caceres2021sparse,haenel2021traversable,berenguer2024floquet,basteiro2024wormhole},
showed that (marginally) relevant tunneling interactions between two such SYK Hamiltonians could induce a phase transition to a finite temperature ``wormhole'' state as the temperature is lowered. The quantum mechanical understanding of this transition is as follows. At high temperatures, when tunneling is unimportant, the state of the coupled system is arbitrarily close to that of two independent decoupled thermal ensembles --- holographically dual to two black holes.
As the temperature is lowered and tunneling becomes important, the thermodynamically preferred state is a different one. This is the state close to the maximally entangled so-called thermo-field double (TFD) state --- holographically dual to a wormhole(WH) connecting the two black holes
\begin{align}
\label{eq:WH-state}
    |\text{WH}\rangle_{\beta} =|\text{TFD}\rangle_{\beta} = \frac{1}{\sqrt{Z_{\beta}}}\sum_n e^{-\frac{\beta}{2}E_{n}} |n,\bar{n}\rangle
\end{align}
with (pure state) density matrix
\begin{align}
    \label{eq:WH-density-matrix}
    \rho_{\text{WH}} = \frac{1}{Z_{\beta}} \sum_{n_1,n_2} e^{-\frac{\beta}{2}(E_{n_1}+E_{n_2})} |n_1,\bar{n}_1\rangle\langle \bar{n}_2,n_2|
\end{align}
Here $Z_{\beta}=\sum e^{-\beta E_n}$ is the partition function of a single SYK model,
and $|\bar{n}\rangle = |\Theta n\rangle$ with $\Theta$ an anti-unitary symmetry of the SYK model (usually CPT).

However theoretically appealing a dual wormhole description of a quantum critical based TFD state is, in actual physical systems quantum critical states are rare.
Moreover, experimentally they are fragile and extremely susceptible to decay to a conventional ordered state, usually a 
superconductor. It is conjectured that in 2D systems this is in fact always the case \cite{metlitskiAreNonFermiliquidsStable2015,abanov2020interplay,chubukov2020interplay}, although 
recent studies have begun to challenge this view ~\cite{zhang2023density}.
The SYK-like quantum critical systems are essentially 0D quantum dots, but 2D extensions can preserve many of its features including the quantum critical (N-AdS$_2$) structure of its groundstate \cite{patel2023universal,esterlis2021large,guo2022large,li2024strange}.  
Some of the expected 2D fragility can be absent accidentally: i.e. 
the simplest SYK models do not preserve time-reversal symmetry and singlet Cooper pairs are not stable enough to form a phase coherent state. 
{More realistic models must also reproduce the quantum critical Eliashberg equations, which can also be obtained purely phenomenologically~\cite{abanov2020interplay,esterlis2025quantumcriticaleliashbergtheory}.}
The most straightforward such model is the Yukawa-SYK model constructed in \cite{esterlis2019cooper}.
This is a quantum critical superconductor with a tunable pairing strength and tunable $T_c$, whose condensate is not controlled by the density of states at the Fermi surface \cite{sheBCSSuperconductivityQuantum2010, she2011observing,esterlis2019cooper,inkof2022quantum,esterlis2025quantumcriticaleliashbergtheory}. 

In this article we address whether
a wormhole state can still arise in tunneling contacts between such more realistic SYK models as Yukawa-SYK that do exhibit superconductivity. 
The notion that superconductivity prevents quantum criticality would argue not: the onset of superconductivity gaps the low-energy spectrum and the continuous quantum critical spectrum out of which the wormhole forms would be absent. As we shall show, the loophole here is that the excitations above the superconducting gap still preserve remnants of the quantum critical state.
   
Furthermore, the superconducting TFD/WH state has a unique signature that can be understood as a generalization of Andreev reflection. The anomalous propagator has a unique revival peak that is holographically dual to an electron tunneling through the wormhole but returning as a coherent hole. As we will show this peak should also be detectable in the Josephson current, where it is interpreted as enhanced Cooper pair tunneling through wormhole traversal. 
{This finding of such a high sensitivity detectable signal of a TFD/wormhole together with the fact that (Y)SYK-models are, on the one hand, now seen as the most promising description of high $T_c$ cuprate (quantum critical) strange metals \cite{patel2023universal,li2024strange}, and, on the other hand, on the road to being realized in e.g. in disordered graphene flakes in high magnetic field ~\cite{chen2018quantum,brzezinska2023engineering,anderson2024magneto,shackleton2024conductance} provides a significant impulse towards observing exotic gravitational phenomena in a table-top laboratory experiment.}

\section{Coupled YSYK models with shared disorder}
The Yukawa-SYK model describes a quantum dot with $N$ flavors of spin-$\frac{1}{2}$ fermions and $M$ flavors of bosons, interacting with each other through a random interaction~\cite{esterlis2019cooper,wang2020solvable,wang2020quantum,classen2021superconductivity,pan2021yukawa,valentinis2023correlation}. When the interaction matrix elements are chosen from the Gaussian Unitary Ensemble(GUE), the ground state is a quantum critical non-Fermi liquid  
with the fermion and boson correlations decaying with a 
conformal power law 
\begin{align}
    G(\tau)  \sim \frac{1}{\abs{\tau}^{2\Delta_f}} \,,
    &~~~~~D(\tau) \sim \frac{1}{\abs{\tau}^{2\Delta_b}} \,.
    \label{eq:scalingsolns}
\end{align}
The scaling dimensions $\Delta_f$ and $\Delta_b$ of the fermions and bosons satisfy the constraint $2\Delta_f + \Delta_b = 1 $. 
They are 
uniquely fixed for a given ratio 
$\kappa=M/N$ ~\cite{wang2020quantum,wang2020solvable,inkof2022thesis}. 
We shall choose $\kappa=1$ with $\Delta_f=0.42\ldots$ throughout. Through the holographic correspondence, the gravity dual of the finite temperature extension of this state is a black hole in AdS$_2$.

\noindent
The system configuration that allows for entangled TFD phase at low temperatures 
is constructed by coupling two identical Yukawa-SYK models, i.e. with the same random coupling constants {\em before disorder averaging}, together with a simple tunneling interaction, described by the following imaginary time action  
\begin{align}
    \label{eq:bareaction}
    S &= S^{(1)}[g_{ijk}] + S^{(2)}[g_{ijk}] + S_c \, ,  \\
    S^{(a)}[g_{ijk}]&= \int_0^{1/T} \!\dd\tau \sum_{i,\sigma} {c^\dagger}^{(a)}_{i\sigma}(\tau) \left(\partial_\tau - \mu\right)c^{(a)}_{i\sigma}(\tau) \nonumber\\ &+ \sum_k \phi^{(a)}_k(\tau)\frac{1}{2}\left(-\partial_\tau^2 + \omega_0^2\right) \phi^{(a)}_k(\tau) \nonumber\\
     &~+ \frac{\sqrt{2}}{N}\int_{0}^{1/T}\!\dd\tau \sum_{ijk,\sigma} g_{ijk}{c^\dagger}^{(a)}_{i\sigma}(\tau)c^{(a)}_{j\sigma}(\tau)\phi^{(a)}_k(\tau) \, ,\nonumber\\
    S_c &= \int_{0}^{1/T} \!\dd\tau \sum_{i,\sigma}\lambda {c^\dagger}^{(1)}_{i\sigma}(\tau)c^{(2)}_{i\sigma}(\tau) + \lambda^* {c^\dagger}^{(2)}_{i\sigma}(\tau)c^{(1)}_{i\sigma}(\tau)  \nonumber
\end{align}
Other tunneling couplings such as boson-boson tunneling $S=\int\dd\tau J\phi^{(1)}_k\phi^{(2)}_k$ or correlated two-fermion tunneling $S=\int\dd\tau g_{ijk}g_{lmk} {c^\dagger}^{(1)}_{i\sigma}(\tau)c^{(2)}_{j\sigma}(\tau) c^{\dagger(1)}_{k\sigma'}c^{(2)}_{l\sigma'} +\text{c.c.}$ are possible. By dimensional analysis, such terms are less relevant in the IR. As a similar study for complex SYK models has shown \cite{sahooTraversableWormholeHawkingPage2020}, with correlated two-fermion tunneling alone, there is still a transition to a TFD/WH state, but it is weaker, at lower temperature and harder to detect. Moreover, such boson-boson and correlated multi-fermion tunnelings are also dynamically generated at subleading order in $\lambda$. We will therefore not consider such couplings here.

By restoring time reversal symmetry,
achieved by choosing the interaction matrix elements $g_{ij,k}$ from the Gaussian Orthogonal Ensemble (GOE) for each $k$, the model has a superconducting ground state instead. 

\noindent
After disorder-averaging over the ensemble, the partition function is dominated 
in the large $N, M$ limit by the saddle point 
solution to the Schwinger-Dyson equations
\begin{align}
    \Sigma_{ab}(\tau,\taup) &= \kappa g^2 D_{ab}(\tau,\taup)G_{ab}(\tau,\taup)\,, \nonumber \\
    \Phi_{ab}(\tau,\taup) &=  -(1-\alpha)\kappa g^2 F_{ab}(\tau,\taup) D_{ab}(\tau,\taup)\,, \nonumber \\
    \Pi_{ab}(\tau,\taup) &= -2 g^2\bigg[
    G_{ab}(\tau,\taup)G_{ba}(\taup,\tau) \nonumber \\
    &\quad -(1-\alpha)F_{ab}(\tau,\taup)\Bar{F}_{ba}(\taup,\tau)
    \bigg]\,.
    \label{eq:CompleteSDeqns}
\end{align}
\begin{align}
\hat{G}(i\omega_n) &= \mqty(i\omega_n-\Sigma_{11} & -\Phi_{11} & -\lambda - \Sigma_{12} & -\Phi_{12} \\ -\bar{\Phi}_{11} & i\omega_n - \tilde{\Sigma}_{11} & -\bar{\Phi}_{12} & \lambda^\ast - \tilde{\Sigma}_{12} \\ -\lambda^\ast - \Sigma_{21} & -\Phi_{21} &i\omega_n-\Sigma_{22} & -\Phi_{22} \\ -\bar{\Phi}_{21} & \lambda - \tilde{\Sigma}_{21} & - \bar{\Phi}_{22} & i\omega_n - \tilde{\Sigma}_{22})^{-\top}\, , \non
\hat{D}(i\nu_n) &= \mqty(\nu_n^2+\omega_0^2-\Pi_{11}(i\nu_n)  & \Pi_{12}(i\nu_n)   \\
    \Pi_{21}(i\nu_n)  & \nu_n^2 + \omega_0^2 - \Pi_{22}(i\nu_n) )^{-\top} \,.
    \nonumber
\end{align}
The parameter $\alpha=1$ for the metallic quantum critical YSYK model, and $\alpha=0$ for the superconducting YSYK model. We have assumed time-translation invariance in addition.
The $4\times 4$ matrix $\hat{G}$ can be expressed in terms of the $2\times2$ matrices in the Nambu basis: $\hat{G} = \mqty(\mathcal{G}_{11} & \mathcal{G}_{12} \\ \mathcal{G}_{21} & \mathcal{G}_{22})$, with each $\mathcal{G}_{ab} = \mqty(G_{ab} & F_{ab} \\ \bar{F}_{ab} & \tilde{G}_{ab})$, where $\tilde{G}(i\omega_n)=-G^\ast(i\omega_n)$ is the Green's function for spin down fermions in the Nambu basis.
In the fermionic matrix equation all self energies are evaluated at the Matsubara frequencies $\Sigma_{ab}(i\omega_n), \Phi_{ab}(i\omega_n)$. 
We will solve the Schwinger-Dyson equations numerically on both the imaginary axis and the the real axis by Anderson iteration methods as in e.g. \cite{maldacenaCommentsSachdevYeKitaevModel2016,esterlis2019cooper,valentinis2023correlation,shankar2023lyapunov,sahooTraversableWormholeHawkingPage2020,pluggeRevivalDynamicsTraversable2020a,lantagne2021superconducting}.
\begin{figure}[t!]
    \centering
    \includegraphics[width=\linewidth]{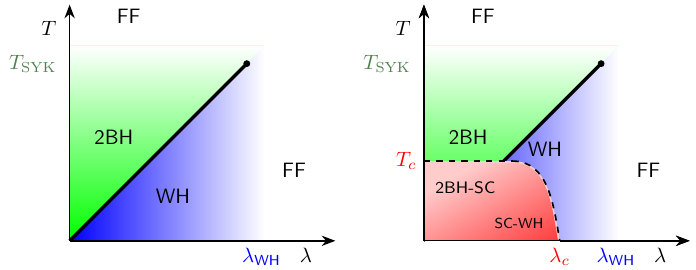}
    \caption{
        Schematic phase diagram of two coupled YSYK systems without (metallic, left) and with superconductivity (right), and 
        $g_{\text{SYK}}/\omega_0^{3/2}\simeq 0.5$ in units of the bare boson mass.
        The configuration of two decoupled, strongly correlated YSYK systems is holographically dual to two thermal black holes (labeled 2BH) . It is separated by a first order phase transition from a low temperature thermofield-double state holographically dual to a wormhole (WH).
        For coupled superconducting YSYK systems, the system first transitions to a superconducting state (2BH-SC) and crosses over to a novel Josephson wormhole state (SC-WH) close to the tunneling strength where superconductivity is suppressed due to pair-breaking tunneling.
        At high temperatures or large tunneling coupling the coupled system is in a free-fermion phase (FF) that smoothly connects.
        }
    \label{fig:schematic-phase-diagram}
\end{figure}

\section{The YSYK-TFD/WH for the metallic state}
We first show that the coupled metallic YSYK model with $\alpha=1$ has a TFD/WH state in the weak tunneling limit, i.e. we first study the system with superconductivity suppressed all the way down to $T=0$.  
Using the mirror symmetry of the action upon permuting the degrees of freedom in the two layers, $c_1 ~\leftrightarrow~ c_2 e^{i\theta}~,~~
\phi_1 \leftrightarrow \phi_2 $
with $\lambda=|\lambda|e^{i\theta}$, the Schwinger-Dyson equations Eq.~\eqref{eq:CompleteSDeqns} can be expressed only in terms of diagonal $G_{11}=G_{22}=G_d$ and off-diagonal propagators $G_{12} = e^{2i\theta}G_{21} = G_{od}$. 
We will make a gauge choice $\lambda$ to be purely real with $\theta = 0$ in what follows.\footnote{In previous work on the complex-SYK model, e.g. Refs.~\cite{sahooTraversableWormholeHawkingPage2020,zhang2021more,chowdhury2022sachdev}, $\lambda$ was chosen to be purely imaginary, such that $G_{12} = -G_{21}$.}

To compare with the superconducting model below, we now choose the SYK coupling to be averaged over an ensemble with zero mean and variance $\langle\langle g_{ijk}g_{ijk}\rangle\rangle_{\text{no sum} \,ijk} = g^2 = 0.25\omega_0^{3}$. Between the three phases of the YSYK model --- the free-fermion phase, the quantum critical phase, and the impurity phase --- {the quantum critical phase sets in before superconductivity for this value } \cite{esterlis2019cooper} (App.A). 
The behavior of the diagonal fermion Green's function for different temperature and tunneling coupling strengths, in relation to the fermionic coherence scale $T_{\text{SYK}}\sim g^{2}/\omega_0^2$, then gives 
the phase diagram of two tunneling coupled metallic YSYKs as shown in Fig.~\ref{fig:schematic-phase-diagram}.
At temperatures much higher than $T_{\text{SYK}}$ one is in a nearly free fermion phase with perturbatively weak SYK interactions, labeled FF.\footnote{We only study the system at charge neutrality and there is no Fermi liquid phase.}
This is continuously connected to the groundstate at large tunnel coupling where $G_{d}(\tau)\sim -\frac{1}{2}\sgn(\tau) e^{-\lambda\abs{\tau}}$. At such large couplings the SYK interactions always remain weak, and one essentially has a set of fermions with an off-diagonal ``mass matrix'' giving rise to the gap above \cite{shankarThesis}.

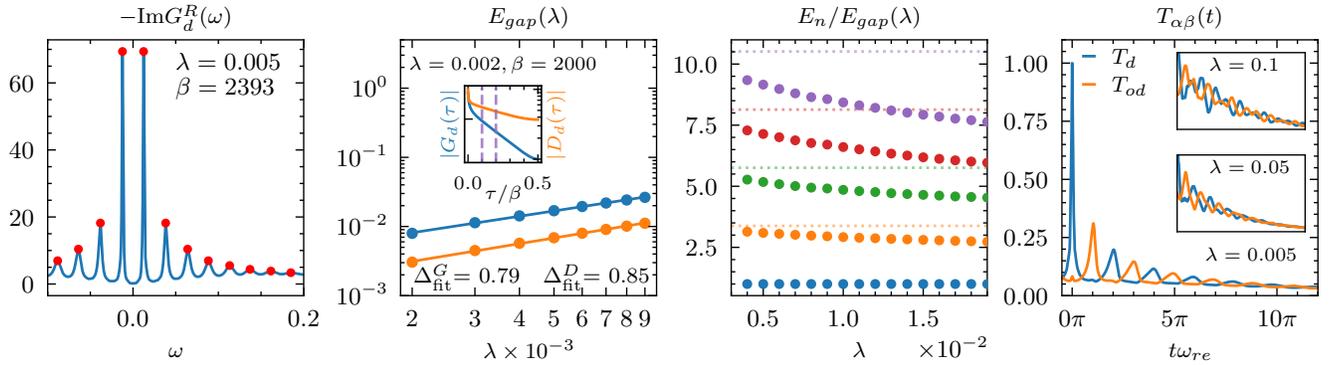
\begin{figure*}
    \centering
    \!
   {\includegraphics[width=\linewidth]{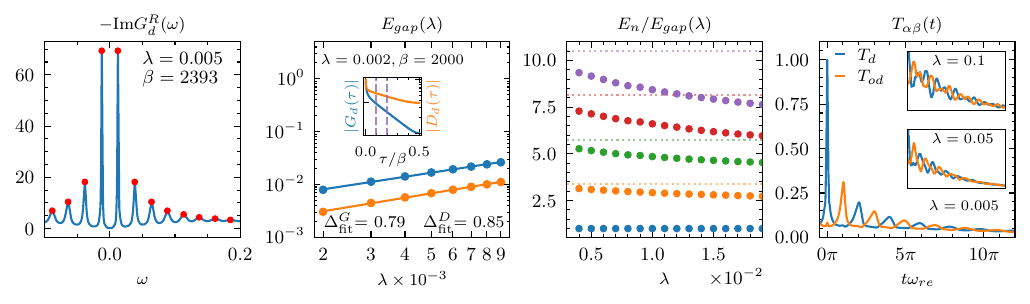}} %
    \setlength{\unitlength}{1mm}
    \begin{picture}(0,0)  
        \put(-66,5){\makebox(0,0)[lt]{\small \textbf{A}}}
        \put(-20,5){\makebox(0,0)[lt]{\small \textbf{B}}}
        \put(25,5){\makebox(0,0)[lt]{\small \textbf{C}}}
        \put(70,5){\makebox(0,0)[lt]{\small \textbf{D}}}
    \end{picture}
    \caption{TFD/WH state signatures in two tunnel-coupled metallic YSYK models with $g/\omega_0^{3/2}=0.5$. {\bf A:} The density of states exhibits the regular linear spacing characteristic of the TFD/WH. {\bf B:} The gap in both the fermionic and bosonic Green's function shows the the expected $\lambda^{\frac{1}{2-2\Delta_f}}$ scaling with $1/(2-2\Delta_f)=0.86$. The inset shows the exponential behavior of the imaginary time Green's functions and the purple dashed lines indicate the region where the fit was performed. 
    {\bf C:} As $\lambda$ decreases, the position of the leading and subleading peaks approach the analytical prediction $E_n/E_{\text{gap}}=(1+\frac{1}{\Delta}n)$ indicated by the dotted lines.
    {\bf D:} Revival oscillations of the transmission amplitude $T_{\alpha\beta}(t) = \frac{2}{\pi}|G_{\alpha\beta}^>(t)|$, with $iG_{\alpha\beta}^>(\omega) = -[1-n_F(\omega)]\text{Im}G^R_{\alpha\beta}(\omega)$. $T_d$ and $T_{od}$ are perfectly out of phase. The characteristic frequency is $\omega_{re}= \frac{p_1}{2\pi}$, where $p_1$ is the average spacing between the peaks in the spectral function.  In the gravitational description, this represents a particle traversing/reemerging from the wormhole.}
    \label{fig:JasperRevivals}
\end{figure*}

At small
tunnel coupling, the interactions do matter.  
For $T<T_{\text{SYK}}$ and $\lambda<\lambda_{\text{WH}}$ 
we find two different phases separated by a first order phase transition {(Fig.\ref{fig:schematic-phase-diagram} left)}~\cite{maldacena2018eternal,maldacenaDivingTraversableWormholes2017,pluggeRevivalDynamicsTraversable2020a,qiCoupledSYKModel2020a}. The high temperature $T>\lambda$ state is the finite temperature extension of each YSYK quantum critical state with negligibly weak tunneling corrections. This is recognizable in a power-law decay of $G_d(\tau)$ according to Eq.~\eqref{eq:scalingsolns} (Fig.~\ref{fig:GreenFunctionPlotsMetal} in Appendix). Its tag 2BH refers to its holographic dual description as two black holes. For $T<\lambda$ there is a new low temperature state. This is the TFD/wormhole state (WH) characterized by a gap and a linearly periodic spectrum $E_n = E_{\text{gap}}(1+\frac{1}{\Delta} n)$ for the lowest lying states \cite{maldacena2018eternal,pluggeRevivalDynamicsTraversable2020a,sahooTraversableWormholeHawkingPage2020}.
The TFD/WH gap is directly visible in the single fermion Green's functions in their exponential decay at large imaginary times $G_d(\tau) \sim e^{-E_{\text{gap}} \tau}$. A characteristic of the wormhole is how
the size of the gap $E_{\text{gap}}$ is controlled by $\lambda$. Its leading functional dependence can be understood from the scaling properties of the tunneling interaction as a perturbation of the quantum critical state \cite{maldacena2018eternal}. In the bilocal formulation of the action this term is 
\begin{align}
\label{eq:tunneling-interaction}
    S_{\text{tunnel}} =\int\!\dd\tau\dd\tau' \lambda G_{12}(\tau,\tau')~,
\end{align}
and hence at the quantum critical point $\lambda \sim \Lambda^{2-2\Delta_f}$ where $\Lambda$ is the RG scale of the theory.
Hence just below the transition we expect the gap to scale as 
\begin{align}
    E_{\text{gap}}[\lambda] \sim \lambda^{\frac{1}{2-2\Delta_f}}. 
    \label{eq:gapscaling}
\end{align}
This scaling is indeed seen by extracting $E_{gap}$ from the logarithmic derivative of the single fermion Green's function $G_d(\tau)$ in the long imaginary time regime for various $\lambda$ at fixed $T\ll \lambda$ in the TFD/WH phase. The boson Green's function  is also gapped and shares the same scaling behaviour as there is only one energy scale that can be constructed using $\lambda$ consistent with the dimensional analysis of Eq.~\eqref{eq:tunneling-interaction}: 
Fig.~\ref{fig:JasperRevivals}B.

The other characteristic of the TFD/WH state --- a linearly spaced low energy spectrum --- can indeed be seen in the single fermion spectral function and gives rise to revival oscillations in the Green's function:
Fig.~\ref{fig:JasperRevivals}A,C\&D. 

\begin{figure}[!t]
    \centering
    \hspace*{-.5in}\includegraphics[width=1.2\linewidth]{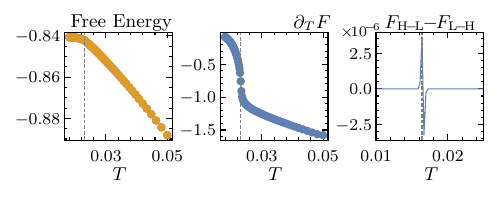}
    \caption{The free energy of two coupled metallic YSYKs with $\lambda = 0.05$, $g/\omega_0^{3/2}=0.5$ 
    shows a first order transition at low $T$ from a 2BH to a WH configuration. (Right) The first order nature also shows up as a hysteresis
    between the solutions obtained from annealing from high to low temperature ($F_{H-L}$), and low to high temperature ($F_{L-H}$). 
    }
    \label{fig:butterflyplotmetal}
\end{figure}

The first order phase transition between 2BH and the TFD/WH states follows from the temperature dependence of the free energy of the system: Fig.~\ref{fig:butterflyplotmetal}. Though not perfectly discontinuous, the derivative of the free energy exhibits a clear jump. Furthermore, while for temperatures away from the critical point, the numerical solution is unique upon annealing both upward and downward in temperature, there is a hysteresis near the phase transition indicating that it is indeed first order. The hysteresis is small and not as sharp as in simpler models in Refs.~\cite{pluggeRevivalDynamicsTraversable2020a,sahooTraversableWormholeHawkingPage2020}, but it is clearly there. 

\section{The Josephson wormhole}
We now show our main result: the TFD/WH state anchored on quantum criticality persists even in the presence of superconductivity.
%
\begin{figure}
    \centering
    \hspace*{-.1in}
    \includegraphics[width=1.05\linewidth]{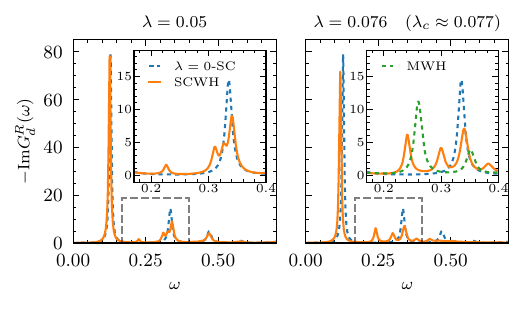}
    \caption{(Left) Diagonal fermionic spectral function in the super\-conducting YSYK model without tunneling (dashed) and with weak tunneling $\lambda = 0.05$ (solid) at low temperature, well below $T_{TFD}$. Compared to the contactless superconductor, the tunneling contacted superconductor shows extra resonances. (Right) The additional resonances can be identified with the metallic TFD/WH state (dashed, green) by tuning towards $\lambda_c$. The superconducting resonance splits into two peaks that disappear in the metallic state and a peak that is a higher wormhole resonance.}
    \label{fig:diffG}
\end{figure}
    
\noindent 
Superconductivity in the YSYK model with $\alpha=0$ at weak coupling $g/\omega_0^{3/2}=0.5$ is known~\cite{esterlis2019cooper,classen2021superconductivity,valentinis2023correlation} to set in just below 
$T_{\rm SYK}$, at which fermionic coherence sets in. Consequently, in the coupled YSYK system at very large $\lambda$, where no quantum critical state occurs, we do not find any evidence of superconductivity, even down to the lowest temperatures that we can reach in our numerics. This can be understood because the non-superconducting state in this case is a conventional gapped free fermion state at charge neutrality. There is no macroscopic density of states at zero energy, and hence no reservoir for Cooper pairs to form or condense.
On the other hand, for $\lambda \ll \lambda_{\text{WH}}$, the superconducting transition temperature 
{$T_c\gg \lambda$} and the superconducting gap $\Delta$ in the coupled model are found to have exactly the same value as they did in the uncoupled model. {Interpolating,} $T_c$ therefore varies with $\lambda$ as in Fig.\ref{fig:schematic-phase-diagram} {(right)} (numerically verified in Fig.\ref{fig:phasebound_lambdamove} in Appendix){, and superconductivity is destroyed when $\lambda$ reaches a critical value $\lambda_c \lesssim \lambda_{WH}$.}

\begin{figure*}[t!]
    \centering
    \includegraphics[width=\linewidth]{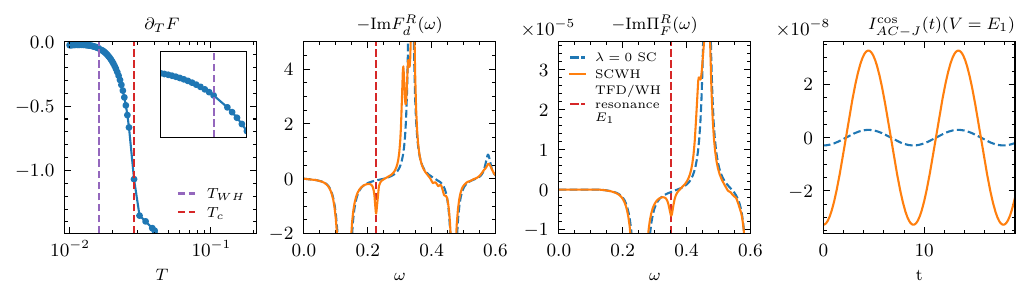}
    \setlength{\unitlength}{1mm}
    \begin{picture}(0,0)  
        \put(-66,5){\makebox(0,0)[lt]{\small \textbf{A}}}
        \put(-20,5){\makebox(0,0)[lt]{\small \textbf{B}}}
        \put(25,5){\makebox(0,0)[lt]{\small \textbf{C}}}
        \put(70,5){\makebox(0,0)[lt]{\small \textbf{D}}}
    \end{picture}
    \caption{
     {\bf A:} Free energy of the coupled superconducting YSYK with $\lambda = 0.05$,  
    $g/\omega_0^{3/2}=0.5$. We observe a second-order transition to the strongly coupled superconducting state at $T_c$, and then at a lower $T<T_c$ a cross-over --- to our numerical accuracy --- from this 2BH-SC to a Josephson wormhole configuration. $T_{\rm WH}$ denotes the temperature where the metallic non-superconducting coupled YSYKs show the first order transition to the TFD/WH state. 
    {\bf B:}
    Andreev-revival peaks characteristic of the TFD/WH state show up in the anomalous Green's function in comparison to the contactless superconductor (blue, dashed). {\bf C:} {This is detectable as a large increase in amplitude Im$\Pi^R_F(\omega)$ in the cosine term of the AC Josephson current ({\bf D}) when the bias voltage is equal to a TFD/WH resonance. }
    }
    \label{fig:FreeEnergySupCondState}
\end{figure*}
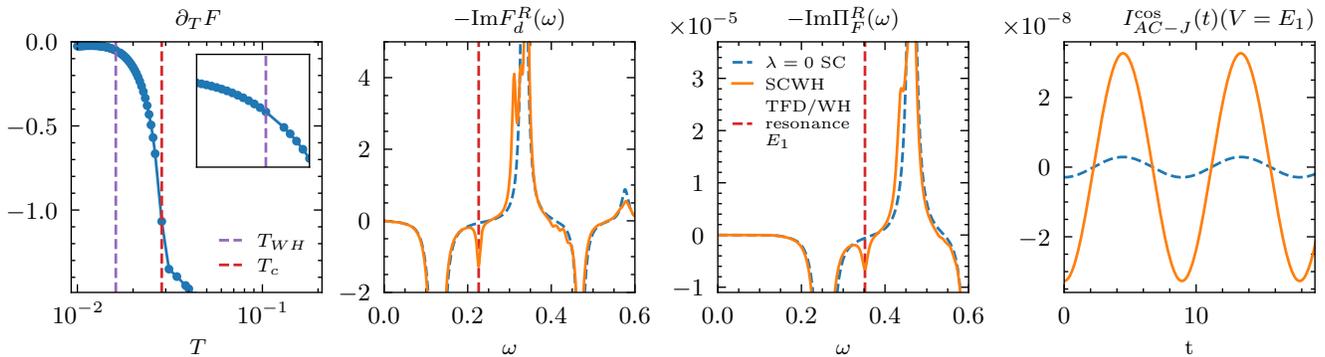
 
For temperatures immediately below $T_c$, the superconducting state --- labeled 2BH-SC --- is nearly identical to two copies of the
superconductor \cite{inkof2022quantum} with macroscopically neglible cross-tunneling corrections between the two subsystems. The diagonal Green's functions are numerically identical to the uncoupled superconducting model at the same temperature,
characterized by a sharp bosonic mode at frequency $\omega_r^{\rm SC}$ and a fermionic spectral function with a superconducting gap $\Delta$ along with higher-order resonances \cite{esterlis2019cooper}. 

Deeper below the TFD/WH transition temperature (as defined for the corresponding non-superconducting system), the superconducting gap persists, but new resonances associated with the TFD/WH state arise. In the bosonic spectrum $\omega_r^{\rm SC}$ splits in two distinct modes {(shown in Fig.~\ref{fig:bosonSpectralFunction} in Appendix.)}, one of which can be identified with $E_{\text{gap}}^{\text{boson}}$ related to the TFD/WH state. In the fermionic spectral function, this is reflected by new peaks and a splitting of the first superconducting resonance peak: Fig.~\ref{fig:diffG}. 
Tuning $\lambda$ to its critical value $\lambda_c$ where superconductivity is lost, the peaks that persist in the non-superconducting state can be identified as TFD/WH resonances. In a small window $\lambda \lesssim \lambda_c$,  these peaks show similar $\lambda$-dependence as  the metallic TFD/WH-state (Fig. \ref{fig:phasebound_lambdamove}B Appendix). {This can be contrasted to conventional subgap Andreev bound states, whose energy decreases quadratically with $\lambda$ for weak Josephson tunneling~\cite{beenakker1991universal}.} In this small region, the fermion spectral function can be thought of as the sum of a superconducting response and the higher order resonances of the TFD/WH state:
\begin{align}
\nonumber
\text{Im} G_d(\omega) \sim  a_{\text{SC}} \text{Im}G_{d,\text{SC}}(\omega) + a_{\text{TFD}}\text{Im}G_{d,\text{TFD}}(\omega)+\ldots
\end{align}
The free energy across the TFD/WH transition, however, shows no discontinuity or hysteresis.
In the presence of superconductivity the first order transition becomes a cross-over and the tunneling coupled superconducting YSYK system has a phase diagram as in the RHS of Fig.~\ref{fig:schematic-phase-diagram}. That the end-point of the critical line is above $T_c$ follows from the fact that $T_c$ decreases with $\lambda$.

These TFD/WH {resonances} are also detectable in the anomalous Green's function $F_d(\omega)$: Fig~\ref{fig:FreeEnergySupCondState}B. 
Similar to 
how the single fermion Green's function is understood as the probability amplitude of detecting an electron at time $t$ given an electron at time $t'$, the anomalous Green's function can be understood as the amplitude of detecting a hole at time $t$ given an electron at time $t'$. This is the perspective how, in a Josephson  set-up, one can view Cooper pair tunneling as Andreev reflection of an electron off the tunneling barrier returning as a hole (at the cost of a Cooper pair absorbed in the condensate). By this viewpoint, the resonance peaks in the anomalous Green's function are Andreev-reflection revivals. 

Such Andreev revivals show up in the AC Josephson current: 
\begin{align}
I_{\text{AC-}J}(t) & = 2\lambda^2 \left[\text{Re}\Pi^R_F(V)\sin(2Vt) +\text{Im}\Pi^R_F(V)\cos(2Vt) \right]\nonumber
\end{align}
with $\Pi_F(\nu_n) = T\sum_n\bar{F}(\omega_m)F(\omega_m-\nu_n)$~\cite{mahan2013many,bruus2004many} the retarded Cooper pair propagator in imaginary time, \mbox{$\omega_m=(2m+1)\pi T$} and \mbox{$\nu_n=2n\pi T$} fermionic and bosonic Matsubara frequencies, and $V$ the applied voltage bias. Specifically, the coefficient of $\cos(2Vt)$ extracts the imaginary part of $\Pi^R_F$ and this can be used not only to identify the higher pairing bound state resonances \cite{esterlis2019cooper}, but also the Andreev-revivals characteristic of the TFD/WH state. 

{
\section{Outlook}
A 2D extension of the Yukawa-SYK model considered here is
the leading theoretical candidate for high-$T_c$ superconductors ~\cite{patel2023universal}. A mesoscopic grain of a cuprate superconductor in its strange metal phase at a temperature $T < v_F/L$ is expected to be a realization of the 0+1D model considered in this article. An AC Josephson current measurement on a device made by coupling two such grains is a good candidate for measuring the new Andreev revival that is described presently --- the BSCCO devices of Refs.~\cite{yu2019high,zhao2019sign,zhao2023time} are a promising avenue.
The existence of a TFD Josephson state holographically dual to wormhole shown here opens a clear way to experimentally test the holographic duality, and observe (quantum) gravitational effects in the laboratory. 
}

\begin{acknowledgments}
\noindent
\section{Acknowledgments}
We thank L. Barbera, C. Beenakker, V. Cheianov, N. Doerfler, I. Jang, J. Schmalian, and especially D. Valentinis and N. Chagnet both for discussions and help with the numerics.
This research was supported in part by the Dutch Research Council (NWO) project 680-91-116 and by the Dutch Research Council/Ministry of Education.
SP thanks C. Beenakker for his kind support during postdoctoral research in Leiden.
The numerical computations were carried out in part on the ALICE-cluster of Leiden University. We are grateful for their help. 
\end{acknowledgments}

\section*{Data availability}
{The data used for the plots generated in this paper can be found at Ref.~\cite{ZenodoUrl}}

\appendix
\section*{Appendix}
\vspace{-3em}
\noindent
\subsection{A. Metallic YSYK TFD/WH revivals in imaginary time Green's functions}
\begin{figure}[h!]
    \centering
    \includegraphics[width=\linewidth]{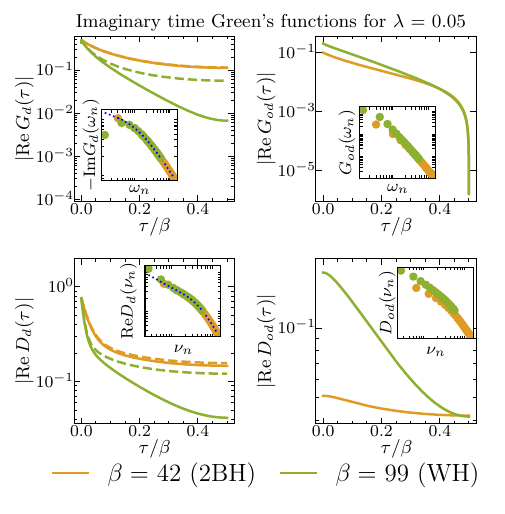}
    \caption{Imaginary time Green's functions at on lowering temperature at fixed $\lambda$ in the metallic state. The dotted lines show the exact numerical solution to contactless model, i.e. when $\lambda = 0$, at the corresponding temperature.  
    The insets show the frequency dependence of the respective Green's functions.}
    \label{fig:GreenFunctionPlotsMetal}
\end{figure}

\noindent
For completeness we show that the TFD/WH gap compared to the finite $T$ quantum spin liquid/2BH state can already be seen in imaginary time as the leading and subleading fall-off in the Green's functions: Fig.\ref{fig:GreenFunctionPlotsMetal} \cite{shankarThesis}.

\noindent
\subsection{B. Superconducting $T_c$ in tunneling coupled YSYKs as a function of $\lambda$}

\begin{figure}[h!]
    \centering
    \includegraphics[width=\linewidth]{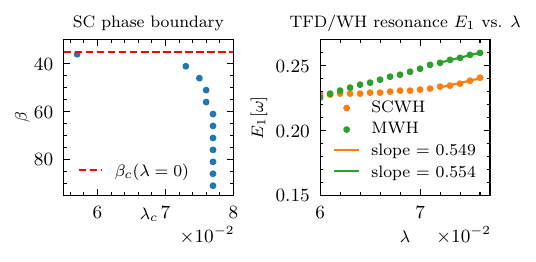}
    \caption{(Left) Numerical results for the superconducting phase boundary justifying the sketch in figure \ref{fig:schematic-phase-diagram}. (Right) $\lambda$-dependence of the TFD/WH resonance in the superconducting model.}
    \label{fig:phasebound_lambdamove}
\end{figure}

\noindent
We have numerically verified the superconducting phase boundary in the tunneling coupled superconducting YSYK models as sketched in Fig.\ref{fig:schematic-phase-diagram}. The critical $\lambda_c$ is defined in these calculations as the value for which the superconducting order parameter $\text{Re}F^R(\omega=0)$ diminishes to below a set threshold: \mbox{Fig. \ref{fig:phasebound_lambdamove}A.}

\noindent
\subsection{C.  $\lambda$-dependence of TFD/WH resonances in the coupled superconducting model}

\noindent
Comparing the TFD/WH resonances as seen in the superconducting model with its metallic counterpart, their dependence is very similar in a small region $\lambda \lesssim\lambda_c$. Outside this region, the resonance location is fixed: Fig. \ref{fig:phasebound_lambdamove}B. 

\noindent
\subsection{D. DC Josephson current}
\begin{figure}[h!]
    \centering
    \includegraphics[width=0.5\textwidth]{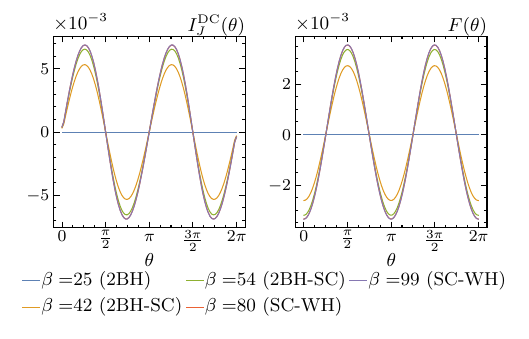}
    \caption{The DC Josephson current and free energy as a function of the phase angle show a small enhancement at temperatures below the TFD/WH transition.}
    \label{fig:enter-label}
\end{figure}

\noindent
In the superconducting state the two tunneling coupled YSYK models form a classic Josephson configuration. We can impose a superconducting phase bias $2\theta$ in units of the superconducting flux quantum between the two sides, and compute its effect on the free energy. This has the following simple form after substituting in $F_{od} = \Phi_{od}=0$: 
\begin{align}
    \mathcal{F}  &= -\frac{1}{\beta}\sum_{\omega_n}\ln\left( f_0(\Sigma,\lambda,\omega_n) + f_2(\Sigma,\lambda,\omega_n)\cos{2\theta}\right) \,,\nonumber \\ 
    f_2 &= 2 \abs{\Phi_d}^2\abs{\lambda + \Sigma_{od}}^2  \, .
    \label{eq:exactFEJosWH_afterphiodzero}
\end{align}
\noindent Eq.~\eqref{eq:exactFEJosWH_afterphiodzero} does not assume weak tunnel coupling,  
and its derivative yields the non-dissipative Josephson current from the Josephson relation $I(\theta) = \partial_\theta {\cal F} =I_c\sin(2\theta)$
\cite{bruus2004many,mahan2013many}. {This technique has also been used to compute Josephson currents in other SYK setups~\cite{lantagne2021superconducting}.}
We can see that $\Sigma_{od}$ acts as a vertex correction to $\lambda$. As a consequence there is a small extra enhancement to the Josephson current once we cross-over to the TFD/WH state.
The DC Josephson current is shown in Fig~\ref{fig:enter-label}. 
The additional boost to the current from the enhancement of $\Sigma_{od}$ in the TFD-state is too small to be noticeable, however, and the fact that the transition is a cross-over also means any derivatives cannot discern this. 

\vspace{-3em}
\noindent
{
\subsection{E. Boson spectral function}
\noindent
Here we show the spectral function for the boson propagator in the Josephson wormhole state. As described in the main text, below the TFD/WH transition temperature, the boson frequency $\omega_r^{SC}$ splits into two. 
The strongest mode can be identified with the pairing channel of the superconductor. By comparing with the YSYK model without superconductivity,the weaker mode at higher frequency can be recognized as a channel associated with the TFD/wormhole state.
\begin{figure}[h!]
    \centering
    \includegraphics[width=\columnwidth]{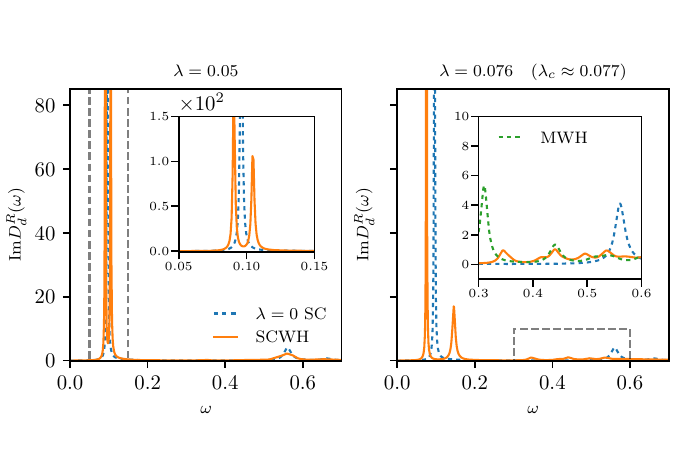}
    \caption{The boson spectral function is characterized by a splitting of $\omega_r^{SC}$ into the pairing channel and a TFD/wormhole channel when below the TFD/WH scale.}
    \label{fig:bosonSpectralFunction}
\end{figure}
}
\vspace{-2em}
{
\subsection{F. Phase diagram of a single YSYK model}
\noindent {Phase diagram of a single YSYK model} with  GOE disorder averaged couplings such that time-reversal symmetry is broken and superconductivity occurs at low $T$: Fig.\ref{fig:phaseDiagSingleYSYK}.
\begin{figure}[h!]
    \centering
    \includegraphics[width=0.9\columnwidth]{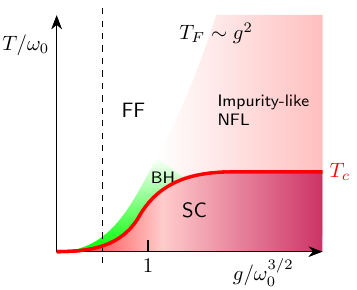}
    \caption{For $T_F\sim g^2/\omega_0^3 < 1$, the high temperature state is characterized by a free fermion phase (FF). Lowering the temperature below $T\lesssim T_F$, the system crosses over to a non-Fermi liquid with a power law self energy, holographically dual to a black hole state (BH). On further lowering temperature, superconductivity sets in at the same characteristic scale $T_c \lesssim T_F$, but always with a lower multiplying factor. In contrast when $g^2/\omega_0^3 > 1$, the fermions freeze into an impurity like regime, where $T_c$ is largely unaffected by the coupling constant.
To ensure the existence of a BH-WH state in a double copy of the YSYK model, we have made the parameter choice indicated by the dashed line in this article.}
    \label{fig:phaseDiagSingleYSYK}
\end{figure}
}

\bibliographystyle{custom-Chagnet}
\bibliography{references}

\end{document}